\documentclass[conference,compsoc]{IEEEtran}

\usepackage[utf8]{inputenc}
\usepackage[T1]{fontenc}

\usepackage[english]{babel}
\usepackage[nocompress]{cite}
\usepackage[pdftex]{graphicx}
\usepackage[cmex10]{amsmath}
\usepackage{amssymb}
\usepackage{algorithmicx}
\usepackage{array}
\usepackage{booktabs}
\usepackage{hyperref}
\usepackage[caption=false,font=footnotesize,labelfont=sf,textfont=sf]{subfig}
\usepackage{dblfloatfix}
\usepackage{url}
\usepackage{tikz}
\usepackage{pgfplots}
\usepackage{pgfplotstable}
\usepackage[draft]{minted}

\IEEEtriggeratref{28}

\tikzstyle{st2b}=[rectangle, minimum height=8pt, minimum width=.4pt, inner sep=0pt, draw]

\setminted{fontsize=\footnotesize}

\definecolor{rwthblue}     {RGB}{  0,  84, 159} 
\definecolor{rwthmagenta}  {RGB}{227,   0, 102} 
\definecolor{rwthyellow}   {RGB}{255, 237,   0} 
\definecolor{rwthpetrol}   {RGB}{  0,  97, 101}
\definecolor{rwthturquoise}{RGB}{  0, 152, 161}
\definecolor{rwthgreen}    {RGB}{ 87, 171,  39}
\definecolor{rwthred}      {RGB}{204,   7,  30}
\definecolor{rwthorange}   {RGB}{246, 168,   0} 
\definecolor{rwthviolet}   {RGB}{ 97,  33,  88} 
\definecolor{rwthpurple}   {RGB}{122, 111, 172}

\pgfplotsset{
    discard if/.style 2 args={
        x filter/.code={
            \edef\tempa{\thisrow{#1}}
            \edef\tempb{#2}
            \ifx\tempa\tempb
                
            \fi
        }
    },
    discard if not/.style 2 args={
        x filter/.code={
            \edef\tempa{\thisrow{#1}}
            \edef\tempb{#2}
            \ifx\tempa\tempb
            \else
                
            \fi
        }
    },
    every tick label/.append style={font=\small},
    /pgf/number format/1000 sep={}
}

\usepackage{xspace}                             
\makeatletter
\DeclareRobustCommand\onedot{\futurelet\@let@token\@onedot}
\def\@onedot{\ifx\@let@token.\else.\null\fi\xspace}
\def\eg{{e.g}\onedot} 
\def\ie{{i.e}\onedot}

\makeatother

\hypersetup{
	pdftitle={Efficient Pattern Matching in Python},
	pdfauthor={Manuel Krebber, Henrik Barthels, Paolo Bientinesi},
	plainpages=false,
	colorlinks=false,
	pdfborder={0 0 0},
	breaklinks=true,
	bookmarksnumbered=true,
	bookmarksopen=true
}

\newcommand{\termset}{\mathcal{T}}
\newcommand{\funcset}{\mathcal{F}}

\newcommand{\varset}{\mathcal{X}}
\newcommand{\groundset}{\mathcal{G}}
\newcommand{\vars}{\mathcal{V}ar}

\newcommand{\substs}{\mathcal{S}ub}

\newcommand{\regvar}[1]{#1}
\newcommand{\plusvar}[1]{{#1}^+}
\newcommand{\starvar}[1]{{#1}^*}

\newif\ifminted
\mintedtrue

\begin{document}

\title{Efficient Pattern Matching in Python}

\author{\IEEEauthorblockN{Manuel Krebber, Henrik Barthels, Paolo Bientinesi}
\IEEEauthorblockA{Aachen Institute for Advanced Study in Computational Engineering Science\\
High-Performance and Automatic Computing Group\\
RWTH Aachen University}}

\maketitle


\begin{abstract}
Pattern matching is a powerful tool for symbolic computations.
Applications include term rewriting systems, as well as the manipulation of symbolic expressions, abstract syntax trees, and XML and JSON data. It also allows for an intuitive description of algorithms in the form of rewrite rules.
We present the open source Python module MatchPy, which offers functionality
and expressiveness similar to the pattern matching in Mathematica.
In particular, it includes syntactic pattern matching, as well as matching
for commutative and/or associative functions, sequence variables, and matching with constraints.
MatchPy uses new and improved algorithms to efficiently find matches for large pattern sets by exploiting similarities between patterns.
The performance of MatchPy is investigated on several real-world problems.
\end{abstract}

\IEEEpeerreviewmaketitle

\section{Introduction}

Pattern matching is a powerful tool which is part of many functional programming
languages as well as computer algebra systems such as Mathematica.
It is useful for many applications including symbolic computation, term simplification,
term rewriting systems, automated theorem proving, code generation, and model checking.
In this paper, we present the Python pattern matching library MatchPy \cite{MatchPy} and its underlying algorithms.

The applications of pattern matching are similar to those of regular expressions, but on symbolic tree structures instead of strings.
Moreover, unlike regular expressions, pattern matching can handle nested expressions up to arbitrary depth.
The goal of pattern matching is to find a match substitution given a subject term and a pattern, which is a term with variables \cite{Baader1998}.
The substitution maps variables in the pattern to replacement terms.
A match is a substitution that, when applied to the pattern, yields the original subject.
As an example consider the subject $f(a)$ and the pattern $f(x)$ where $f$ is a function symbol, $a$ is a constant and $x$ is a variable.
Then the substitution $\sigma = \{ x \mapsto a \}$ is a match because $\sigma(f(x)) = f(a)$.
The most basic form of pattern matching where functions are neither associative nor commutative is called syntactic matching.



In addition to regular variables which can match a single term, MatchpPy supports
sequence variables that can match a sequence of terms.
For example, if $\plusvar{x}$ is such a sequence variable, then $f(a, b)$ matches the pattern $f(\plusvar{x})$ with the substitution $\{ x \mapsto (a, b) \}$, whereas $f(\regvar{x})$ does not match.
MatchPy enables further control over what variables can match by supporting symbol variables,
\ie variables that only match a specific user-defined class of symbols instead of any term.
Furthermore, variables can have a default value, as for instance in the pattern $x{:}1 \cdot y$. Here, $x{:}1$ denotes that if
$x$ does not match anything else, then it has a default value of $1$. This
pattern matches both $a$ and $a \cdot b$. For $a$, the match is $\{ x \mapsto 1, y \mapsto a \}$, whereas for the subject $a \cdot b$, it is $\{ x \mapsto a, y \mapsto b \}$.

In practice, functions commonly have properties such as commutativity and associativity.
These properties affect the pattern matching, \eg $f(a, b)$ and $f(b, a)$ match iff $f$ is commutative.
To our knowledge, no existing work covers pattern matching with function symbols which are either commutative or associative but not both at the same time.
However, several common functions possess such properties,  \eg matrix multiplication and arithmetic mean.

Among the existing systems, Mathematica~\cite{Mathematica} arguably offers the
most powerful pattern matching.
Patterns are used widely in Mathematica, \eg in function definitions or for
manipulating terms.
Mathematica offers support for associativity, commutativity, sequence variables, optional variables, symbol variables, constraints, alternative patterns, and arbitrarily repeated patterns.
MatchPy aims to replicate most of this functionality.

Pattern matching forms the basis of term rewriting systems (TRS), where it is necessary to determine which rewrite rules can be applied. Since a TRS maps input expressions to output expressions, they can be seen as algorithms or programs operation on symbolic expressions. As a sufficiently powerful TRS can be Turing complete, they can be used as a programming language \cite{Kirchner2001}.

In many applications, a fixed pattern set is matched repeatedly against different subjects.
The simultaneous matching of multiple patterns is called many-to-one matching, as opposed to one-to-one matching.
Many-to-one matching can provide a significant speedup over one-to-one matching by exploiting similarities between patterns.
While there has been research on many-to-one matching for some of MatchPy's features, previously no many-to-one algorithms existed for pattern matching as expressive as the one offered by MatchPy (and Mathematica). We generalize existing algorithms to support associativity, commutativity, sequence/optional/symbol variables, and constraints.



\section{Related Work}

While some basic forms of pattern matching are common in functional programming languages, support for pattern matching in popular imperative programming languages is fairly rare.
Most of the existing pattern matching libraries for Python only support syntactic patterns.
Among them are several that allow for functional-style syntactic pattern matching for native data structures, \eg MacroPy \cite{MacroPy}, patterns \cite{patterns} or PyPatt \cite{PyPatt}.
The pattern matching in SymPy \cite{SymPy} can work with associative and
commutative functions, but it is limited to finding a single match---to enable
further processing, it is often
useful to find all possible matches for a pattern---and there are no algorithms for many-to-one matching.
Furthermore, SymPy does not support sequence variables and is limited to a predefined set of mathematical operations.

Previous research on many-to-one matching either focused on syntactic matching (functions with fixed arity and no special properties) \cite{Christian1993,Graef1991,Nedjah1997}
or AC matching (variadic, associative and commutative functions) \cite{Kounalis1991,Bachmair1993,Lugiez1994,Bachmair1995,Eker1995,Kirchner2001}. This research did not consider function symbols which only have some of those properties, nor sequence variables. The work of Kutsia does include sequence variables, but the focus is on theoretical aspects of one-to-one matching \cite{Kutsia2006,Kutsia2007}.



Even though Mathematica has powerful pattern matching features, it has some drawbacks. The
possibilities to access Mathematica's features from other programming languages is rather limited;
writing large programs in Mathematica can be cumbersome and slow; furthermore,
Mathematica is a commercial and proprietary product. Instead, it is desirable to have a free and
open source pattern matching implementation that also enables other researchers to use and extend it.

Mathics \cite{Mathics} is an open source computer algebra system written in Python that aims to replicate the syntax and functionality of Mathematica.
Unfortunately,  currently there is little active development in this project.

\section{Preliminaries}

The notation and definitions used are based on what is used in term rewriting systems literature \cite{Dershowitz1990,Baader1998,Klop2001}.
Pattern matching works on terms that consist of function symbols $\funcset$ and variables $\varset$.
The function symbol set is composed of function symbols with different arities, \ie $\funcset = \bigcup_{n\geq0} \funcset_n$ where $\funcset_n$ contains symbols with arity $n$.
The function symbols can either have a fixed arity (\ie they only occur in one
$\funcset_n$) or be \emph{variadic} (\ie occur in all $\funcset_n$ for $n \geq n_0$ and some fixed $n_0$).
Specifically, $\funcset_0$ contains all constant symbols.
The set of all terms $\termset(\funcset, \varset)$ is the smallest set such that
\begin{itemize}
\item $\varset \subseteq \termset(\funcset, \varset)$ and
\item for all $n \geq 0$, all $f \in \funcset_n$, and all $t_1, \dots t_n \in \termset(\funcset, \varset)$ we have $f(t_1, \dots, t_n) \in \termset(\funcset, \varset)$.
\end{itemize}
Whenever the actual symbols and variables are not important, we use $\termset$ instead of $\termset(\funcset, \varset)$.
Terms in $\varset \cup \funcset_0$ are called \emph{atomic terms}, all others are \emph{compound terms}.
The set $\termset(\funcset, \emptyset) \subseteq \termset(\funcset, \varset)$ is the set of \emph{ground terms}. We usually use $\groundset(\termset)$ or simply $\groundset$ to denote it.
Since ground terms are variable-free, they are also called
\emph{constant terms}.
The set of variables occurring in a term $t$ is denoted by $\vars(t)$.
A pattern $t$ is called \emph{linear} if every variable in $\vars(t)$ occurs at most once.

In the following, we usually use $f$, $g$, and $h$ as function symbols and $a$, $b$, and $c$ as
constant symbols.
Common mathematical functions such as $+$ and $\times$ are usually written in infix notation, and we omit the parenthesis if unnecessary, \ie we write $a + b$ instead of $+(a, b)$.

A substitution is a function $\sigma\colon \varset \rightarrow \termset$.
We extend the substitution to a function $\sigma\colon \termset \rightarrow \termset$ over all terms by using $\sigma(f(t_1, \dots, t_n)) = f(\sigma(t_1), \dots, \sigma(t_n))$ for every $n \geq 0$, all $f \in \funcset_n$ and all $t_1, \dots t_n \in \termset$.
We often write substitutions as $\sigma = \{ x_1 \mapsto t_1, \dots, x_n \mapsto t_n \}$ where $\{x_1, \dots, x_n\} = Dom(\sigma)$ are the variables \emph{instantiated} by the substitution.
The set of all substitutions is called $\substs(\termset(\funcset, \varset))$ or simply $\substs$.

A pattern term $t \in \termset$ \emph{matches} a subject term $s \in \groundset$, iff there exists a substitution $\sigma$ such that $\sigma(t) = s$.
Such a substitution is also called a \emph{match}.
The objective of pattern matching is to find such match substitutions.

We extend this basic definition of syntactic pattern matching to support associativity, commutativity, and one-identity.
If a function symbol $f$ is associative, \ie $f(t_1, f(t_2, t_3)) = f(f(t_1, t_2), t_3)$ holds for all $t_1, t_2, t_3 \in \termset$,
we always use the equivalent flattened form $f(t_1, t_2, t_3)$.
If a function symbol $f$ is commutative, \ie $f(t_1, t_2) = f(t_2, t_1)$ holds for all $t_1, t_2 \in \termset$,
we always use the equivalent form $f(t_1, t_2)$ with lexicographically sorted arguments (assuming $t_1 \leq t_2$).
Finally, if a function symbol $f$ is equivalent to the identity function with a single argument,
\ie $f(t_1) = t_1$ holds for all $t_1 \in \termset$, we always write $t_1$
instead.
We use $\funcset_a$, $\funcset_c$, and  $\funcset_I$ to denote
the set of all associative functions,
the set of all commutative functions.
and the set of identity functions, respectively.
Finally, we use $=_{ACI}$ to denote equivalence of terms under associativity, commutativity, and one-identity and consider $\sigma$ to be a match if there exist $t' \in \termset$ and $s' \in \groundset$ such that $\sigma(t') = s'$ and $t =_{ACI} t'$ and $s =_{ACI} s'$.

In contrast to regular variables, sequence variables match a sequence of terms instead of a single term.
Sequence variables are denoted analogously to the notation used in regular expressions:
$x^*$ matches any number of terms including the empty sequence, $x^+$ requires at least one term to match.\footnote{Corresponding to \texttt{x\_\_\_} and \texttt{x\_\_} in Mathematica, respectively.}
The definition of a substitution is extended to allow sequences of terms as well: $\sigma\colon \varset \rightarrow \termset^*$.
When applying a substitution, the replacement of a sequence variable is integrated into the sequence of function symbol arguments: $\sigma(f(a, x^*, b)) = f(a, c, d, b)$ for $\sigma = \{ x^* \mapsto (c, d) \}$.
Patterns with sequence variables may yield multiple matches.

Matching sequence variables is equivalent to matching regular variables within associative functions.
As an example, consider $f(a, x^+)$ and $f_A(a, x)$ where $f_A$ is associative.
For the subjects $f(a, b, c)$ and $f_A(a, b, c)$, substitutions are $\{ x \mapsto (b, c) \}$ and $\{ x \mapsto f_A(b, c) \}$, respectively.

MatchPy also supports constraint predicates to further filter matches.
A constraint is a function $\varphi\colon \substs \mapsto \{ 0, 1 \}$.
We say a match $\sigma$ is \emph{valid} for a pattern with constraint $\varphi$ iff $\varphi(\sigma) = 1$.
Symbol variables can be used to match only specific symbol classes, \ie a specific subset $C \subseteq \funcset_0$.
Furthermore, matching a variable can be made optional by providing a default value for when it does not match.
This is useful to reduce the number of patterns.
As an example, $a \cdot x + b$ can cover a linear equation with $a$ defaulting to $1$ and $b$ to $0$. Then the subject $x$ can be matched with $\{ a \mapsto 1, b \mapsto 0, x \mapsto x \}$.

\section{Methods/Optimizations}
\subsection{One-to-one}


\subsubsection{Associativity/Sequence variables%
  \label{associativity-sequence-variables}%
}

Associativity enables arbitrary grouping of arguments for matching:
For example, $1 + a + b$ matches $1 + \regvar{x}$ with $\{ x \mapsto a + b \}$ because we can group the arguments as $1 + (a + b)$.
When regular variables are arguments of an associative function, they behave like sequence variables.
Both can result in multiple distinct matches for a single pattern.
To enumerate all matches, we need to backtrack while matching for every choice that we make for such variables.
To accomplish this, we use Python generators to return a value and be able to resume at the same state when backtracking.
Associative matching is NP-complete \cite{Benanav1987}.

\subsubsection{Commutativity%
  \label{commutativity}%
}

Matching commutative terms is difficult because matches need to be found independently of the argument order.
Commutative matching has been shown to be NP-complete \cite{Benanav1987}.
It is possible to find all matches by matching all permutations of the subjects arguments against all permutations of the pattern arguments.
If $n$ is the number of subject arguments, and $m$ is the number of pattern arguments,
this naive approach leads to testing a total of $n!m!$ combinations,
and it is likely that most of them either do not match or yield redundant matches.
To address this challenge, we interpret the arguments as a multiset, \ie an orderless collection that allows repetition of elements.

Additionally, we use the following sequence of steps for matching the subterms of a commutative term:
\begin{enumerate}
    \item Constant arguments
    \item Matched variables, \ie variables that already have a value assigned in the current substitution
    \item Non-variable arguments
    \item Go back to 2 if there are new matched variables.
    \item Regular variables
    \item Sequence variables
\end{enumerate}

The idea behind this sequence of steps is to reduce the size of the search space as quickly as possible, starting with the simplest steps. As an example, matching constant arguments is equivalent to testing multiset membership, which can be done very efficiently. Each step reduces the search space for successive steps. If one step finds no match, the remaining steps do not have to be performed.
Note that steps 3, 5 and 6 can yield multiple matches and backtracking is employed to check every combination.
Since step 6 is the most involved, it is described in more detail in the next section.

\subsubsection{Sequence Variables in Commutative Functions%
  \label{sequence-variables-in-commutative-functions}%
}

The distribution of $n$ subjects subterms onto $m$ sequence variables within a
commutative function symbol can yield up to $m^n$ distinct solutions.
Enumerating all of the solutions is accomplished by generating and solving several linear Diophantine equations.
As an example, lets assume we want to match $f(a, b, b, b)$ with $f(\starvar{x}, \plusvar{y}, \plusvar{y})$ where $f$ is commutative.
This means that the possible distributions are given by the non-negative integer solutions of these equations:
\begin{eqnarray*}
1 &=& x_a + 2 y_a \\
3 &=& x_b + 2 y_b
\end{eqnarray*}
$x_a$ determines how many times $a$ is included in the substitution for $x$.
Because $y$ requires at least one term, we have the additional constraint $y_a + y_b \geq 1$.
The only possible solution $x_a = x_b = y_b = 1 \wedge y_a = 0$ corresponds to the match substitution $\{ x \mapsto (a, b), y \mapsto (b) \}$.

Extensive research has been done on solving linear Diophantine equations and linear Diophantine
equation systems \cite{Weinstock1960,Bond1967,Lambert1988,Clausen1989,Aardal2000}.
In our case, the equations are actually independent except for the additional constraints for plus variables.
Furthermore, only the non-negative solutions are of interest in this case, and they can be found more easily.
We use an adaptation of the algorithm used in SymPy which recursively reduces
a linear Diophantine equation to a set of equations of the form $ax + by = d$, which can be solved efficiently with the Extended Euclidian algorithm \cite{Menezes1996}.
The solutions are then combined into a solution for the original equation.

Since the coefficients in the equations correspond to the multiplicity of sequence variables,
they are likely very small.
Similarly, the number of variables in the equations is usually small as they map to sequence variables.
The constant is the multiplicity of a subject term and hence also usually small.
Overall, the number of distinct equations that are solved is small. Thus, by caching solutions, the impact of matching sequence variables on the overall run time can be reduced.

\subsection{Many-to-one}

MatchPy includes several many-to-one matching approaches: A deterministic discrimination net that only works for syntactic patterns, a generic many-to-one matcher and a code generator based on the many-to-one matcher.
A discrimination net is a data structure similar to a decision tree or a finite automaton \cite{Christian1993,Graef1991,Nedjah1997}.
Generally, the many-to-one matchers enable matching multiple patterns against a single subject much faster than matching each pattern individually using the one-to-one matching.
The generic matcher utilizes a generalized form of non-deterministic discrimination nets that support sequence variables and associative function symbols.
Furthermore, as described in the next section, it can also match commutative terms.

\begin{figure}
    \centering
    \begin{tikzpicture}[
        grow=right,
        sloped,
        thick,
        level 2/.style={sibling distance=15mm},
        level 3/.style={sibling distance=10mm},
        level distance=15mm,
    ]
        \node[st2b] {}
            child {
                node[st2b] {}
                child {
                    node[st2b] {}
                    child {
                        node[st2b] {}
                        node [label={[right,align=left]\texttt{[1]}}] {}
                        edge from parent
                        node[above] {\texttt{]}}
                    }
                    child {
                        node[st2b] {}
                        child {
                            node[st2b] {}
                            node [label={[right,align=left]\texttt{[1, $\starvar{x}$]}}] {}
                            edge from parent
                            node[above] {\texttt{]}}
                        }
                        edge from parent
                        node[above] {\texttt{$\starvar{x}$}}
                    }
                    edge from parent
                    node[above] {\texttt{1}}
                }
                child {
                    node[st2b] {}
                    child {
                        node[st2b] {}
                        child {
                            node[st2b] {}
                            node [label={[right,align=left]\texttt{[$\regvar{y}$, 0]}}] {}
                            edge from parent
                            node[above] {\texttt{]}}
                        }
                        edge from parent
                        node[above] {\texttt{0}}
                    }
                    edge from parent
                    node[above] {\texttt{$\regvar{y}$}}
                }
                edge from parent
                node[above] {\texttt{[}}
            };
    \end{tikzpicture}
    \caption{Example Discrimination Net. \label{fig:dn}}
\end{figure}
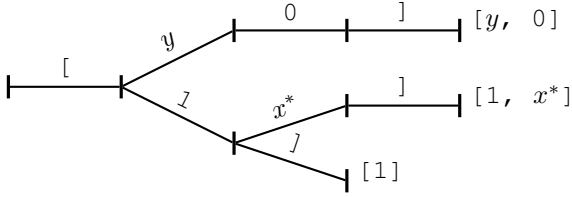

In Figure \ref{fig:dn}, an example for a non-deterministic discrimination net is shown.
It contains three patterns that match Python lists:
One matches the list that consists of a single 1, the second one matches a list with exactly two elements
where the last element is 0, and the third pattern matches any list where the first element is 1.
Note that these patterns can also match nested lists, \eg the second pattern would also match \texttt{{[}{[}2, 1{]}, 0{]}}.

Matching starts at the root and proceeds along the transitions.
Simultaneously, the subject is traversed in preorder and each symbol is checked against the transitions.
Only transitions matching the current subterm can be used.
Once a final state is reached, its label gives a list of matching patterns.
For non-deterministic discrimination nets, all possibilities need to be explored via backtracking.
The discrimination net allows to reduce the matching costs, because common parts of different patterns only need to be matched once.
For non-matching transitions, their whole subtree is pruned and all the patterns are excluded at once, further reducing the match cost.

In Figure \ref{fig:dn}, for the subject \texttt{{[}1, 0{]}}, there are two paths and therefore two matching patterns:
\texttt{{[}$\regvar{y}$, 0{]}} matches with $\{ y \mapsto 1 \}$ and \texttt{{[}1, $\starvar{x}${]}} matches with $\{ x \mapsto 0 \}$.
Both the \texttt{y}-transition and the \texttt{1}-transition can be used in the second state to match a \texttt{1}.

Compared to existing discrimination net variants, we added transitions for the end of a compound term to support variadic functions.
Furthermore, we added support for both associative function symbols and sequence variables.
Finally, our discrimination net supports transitions restricted to symbol classes, \ie symbol variables.
We decided to use a non-deterministic discrimination net instead of a deterministic one, since the number of states of the latter grows exponentially with the number of patterns.
While the deterministic discrimination net also has support for sequence variables, in practice the net became too large to use with just a dozen patterns.

\subsubsection{Commutative Many-to-one Matching%
\label{commutative-many-to-one-matching}%
}

Many-to-one matching for commutative terms is more involved.
We use a nested commutative matcher which in turn uses another generic many-to-one matcher to match the subterms.
Our approach is similar to the one used by Bachmair and Kirchner in their respective works \cite{Bachmair1995,Kirchner2001}.
We match all the subterms of the commutative function in the subject with a many-to-one matcher constructed from the
subpatterns of the commutative function in the pattern (except for sequence variables, which are handled separately).
The resulting matches form a bipartite graph, where one set of nodes consists of the subject subterms and the other contains all the pattern subterms.
Two nodes are connected by an edge iff the pattern matches the subject.
Such an edge is also labeled with the match substitution(s).
Finding an overall match is then accomplished by finding a maximum matching in this graph.
However, for the matching to be valid, all the substitutions on its edges must be compatible,
\ie they cannot have contradicting replacements for the same variable.
We use the Hopcroft-Karp algorithm \cite{Hopcroft1973} to find an initial maximum matching.
However, since we are also interested in all matches and the initial matching might have incompatible substitutions,
we use the algorithm described by Uno, Fukuda and Matsui \cite{Fukuda1994,Uno1997} to enumerate all maximum matchings.

\begin{figure}[]\noindent\makebox[\columnwidth][c]{\includegraphics[width=\columnwidth]{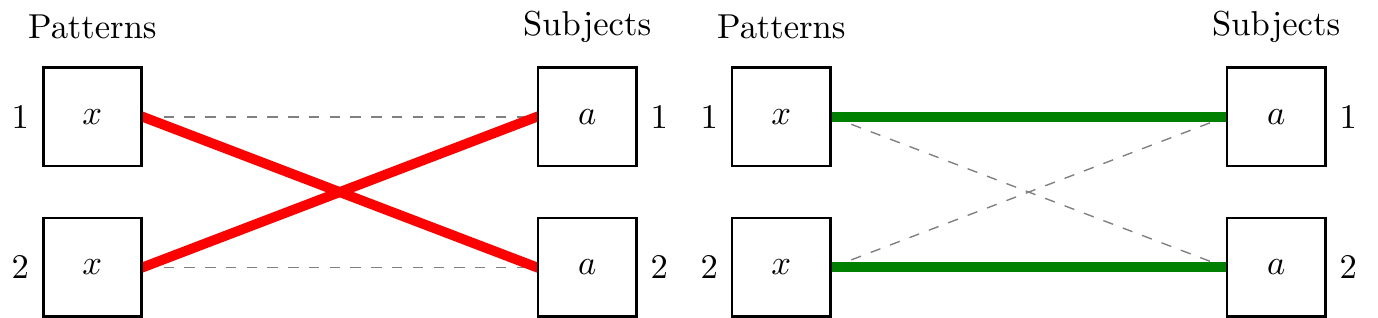}}
\caption{Example for Order in Bipartite Graph. \label{fig:bipartite2}}
\end{figure}

To avoid yielding redundant matches, we extended the bipartite graph by introducing a total order over its two node sets.
This enables determining whether the edges of a matching maintain the order induced by the subjects or whether some of the edges \textquotedbl{}cross\textquotedbl{}.
Formally, for all edge pairs $(p, s), (p', s') \in M$ we require $(s \equiv s' \wedge p > p') \implies s > s'$
to hold where $M$ is the matching, $s, s'$ are subjects, and $p, p'$ are patterns.
An example of this is given in Figure \ref{fig:bipartite2}.
The order of the nodes is indicated by the numbers next to them.
The only two maximum matchings for this particular match graph are displayed.
In the left matching, the edges with the same subject cross and hence this matching is discarded.
The other matching is used because it maintains the order.
This ensures that only unique matches are yielded.
Once a matching for the subpatterns is obtained, the remaining subject arguments are
distributed to sequence variables in the same way as for one-to-one matching.

\subsection{Code Generation}

Both the one-to-one and many-to-one matching algorithm use some in-memory representation of the patterns at run time.
This allows for the dynamic construction of the pattern set at run time.
For fixed pattern sets, however, many-to-matching can be sped up even further by generating Python code, similar to how parser generators generate code that parses a given grammar.
This is done by converting the discrimination net structure of the many-to-one matcher to code.
While this code generation is expensive and yields large code files,
the resulting code offers a significant speedup over regular many-to-one matching.

\section{Experimental Results}

We perform multiple experiments to evaluate the performance of different pattern matching strategies.
The applications comprise linear algebra expressions, Python source code transformation,
converting logic formulas to algebraic normal form (ANF), and symbolic integration.
The evaluated pattern matching methods are one-to-one matching, many-to-one matching,
code generation, and parallelized one-to-one matching.
In every experiment, a single subject is matched against a fixed pattern set.
Since all methods can be parallelized over multiple subjects by using standard methods
(\eg \texttt{multiprocessing} or dask), the experiments do not cover this.
We do not consider the singular setup time for many-to-one matching or code generation
here, because previous experiments have shown that the setup time is easily amortized \cite{Krebber2017,thesis}.

All times are measured on a server machine with 2 Intel Xeon E5-2670 v2 2.5GHz CPUs with 10 cores each and 64GB of RAM running Cent OS 6.8 and Python 3.6.2.
The parallelized matching uses all 20 cores using the \texttt{multiprocessing} module.
Because of the global interpreter lock (GIL) that prevents multithreading within Python code itself, this uses process based parallelism.
We use a pool of processes and a queue for every subject-pattern pair.
Since the pattern set is fixed, the patterns are available directly in the code.
The subjects are considered to be dynamic and dependent on the use of the application. Hence, they are serialized to be sent to the worker processes.

\subsection{Linear Algebra}

\begin{table}
    \centering
    \caption{Linear Algebra Operations}
    \renewcommand{\arraystretch}{1.2}
    \begin{tabular}{l c c p{1.5cm}}
        \toprule
        \textbf{Operation} & \textbf{Symbol} & \textbf{Arity} & \textbf{Properties} \\
        \midrule
        Multiplication & $\times$ & variadic & associative \\
        Addition & $+$ & variadic & associative,\newline commutative \\
        Transposition & ${}^T$ & unary & \\
        Inversion & ${}^{-1}$ & unary & \\
        Inversion and Transposition & ${}^{-T}$ & unary & \\
        \bottomrule
    \end{tabular}
\label{tbl:laop}
\end{table}

BLAS is a collection of optimized routines that can compute specific linear algebra operations efficiently \cite{Lawson1979,Dongarra1988,Dongarra1990}.
As an example, assume we want to match all subexpressions of a linear algebra expression which can be computed by the \href{https://software.intel.com/en-us/node/468494}{?TRMM} BLAS routine.
These have the form $\alpha \times op(A)  \times B$ or $\alpha  \times B  \times op(A)$ where
$op(A)$ is either the identity function or transposition, and $A$ is a triangular matrix.
For this example, we leave out all variants where $\alpha \neq 1$.

In order to model the linear algebra expressions, we use the operations shown in Table \ref{tbl:laop}.
In addition, we have special symbol subclasses for scalars, vectors, and matrices and use symbol variables to match them.
Matrices also have a set of properties, \eg they can be triangular, symmetric, square, etc.
We use constraints to check the properties of the matched symbols.
Finally, sequence variables are used to capture the remaining operands of multiplications or additions.

In total, we have about 200 patterns with about 60 being patterns for sums and 140 for products.
A lot of these patterns only differ in terms of constraints, e.g. there are ten distinct patterns matching $A \times B$ with different constraints on the two matrices.
As subjects, we use about 140 terms representing different linear algebra problems.


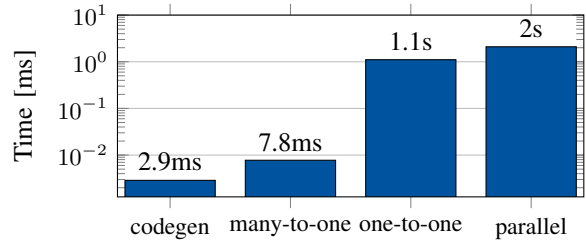
\begin{figure}[h]
    \centering
    \begin{tikzpicture}
        \begin{axis}[
            ybar,
            height=4cm,
            width=\columnwidth,
            ylabel={Time [ms]},
            symbolic x coords={codegen,many-to-one,one-to-one,parallel},
            xtick=data,
            nodes near coords,
            ymin=0,
            ymax=10,
            bar width=1.2cm,
            x=1.6cm,
            enlarge x limits={abs=0.7cm},
            xticklabel style = {font=\small},
            ymode=log,
            log origin=infty,
            point meta=explicit symbolic,
            ymajorgrids,
        ]
            \addplot[fill=rwthblue] plot coordinates {
                (codegen,       0.002877) [2.9ms]
                (many-to-one,   0.007757) [7.8ms]
                (one-to-one,    1.106984) [1.1s]
                (parallel,      2.093992) [2s]
            };
        \end{axis}
    \end{tikzpicture}
    \vspace{-\baselineskip}
    \caption{Total times for \texttt{LinAlg}}\label{fig:linnea-times}
\end{figure}

The results for those \texttt{LinAlg} problems are shown in Figure \ref{fig:linnea-times}.
Both many-to-one matching and the generated code are significantly faster than sequential and parallel one-to-one matching.
Both are about two orders of magnitude faster than one-to-one matching.
The generated code is about 2.7 times faster than the many-to-one matcher.
The parallelized one-to-one matching is slower than the regular one-to-one matching.
This is caused by the overhead of the process-based parallelism which requires the subjects to be serialized to be passed between processes.

\subsection{Abstract Syntax Trees}

Python includes a tool to convert code from Python 2 to Python 3.
It is part of the standard library package \texttt{lib2to3} which has a collection of \textquotedbl{}fixers\textquotedbl{} that each convert one of the incompatible cases.
To find matching parts of the code, those fixers use pattern matching on the abstract syntax tree (AST).
Such an AST can be represented in the MatchPy data structures.
We converted some of the patterns used by \texttt{lib2to3} both to demonstrate the generality of MatchPy and to evaluate the performance of many-to-one matching.
Because the fixers are applied one after another and can modify the AST after each match,
it would be difficult to use many-to-one matching for \texttt{lib2to3} in practice.

\ifminted
The following is an example of such a pattern:
\begin{minted}{python}
power< 'isinstance'
    trailer< '(' arglist< any ','
        atom< '('
            args=testlist_gexp< any+ >
')' > > ')' > >
\end{minted}

It matches an \texttt{isinstance} expression with a tuple as second argument.
Its tree structure is illustrated in Figure \ref{fig:ast}.
The corresponding fixer cleans up duplications generated by previous fixers.
For example, \mintinline{python}{isinstance(x, (int, long))}
would be converted by another fixer into \mintinline{python}{isinstance(x, (int, int))}
which in turn is then simplified to \mintinline{python}{isinstance(x, int)} by this fixer.
\begin{figure}[]\noindent\makebox[\columnwidth][c]{\includegraphics[scale=0.80]{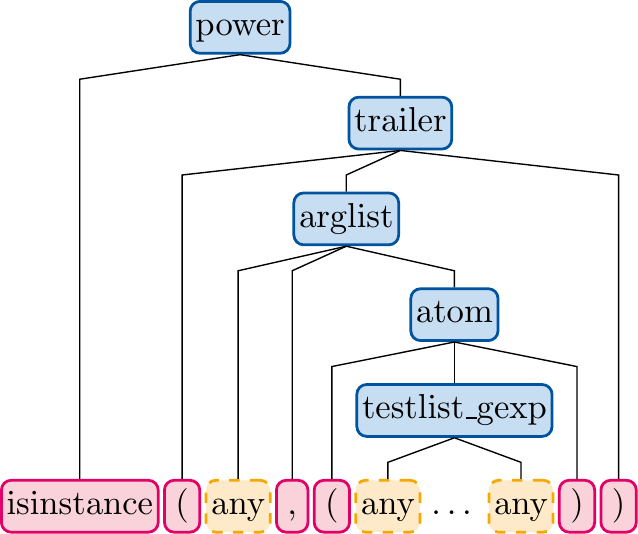}}
\caption{AST of the \texttt{isinstance} pattern. \label{fig:ast}}
\end{figure}

Out of the original 46 patterns, 36 could be converted to MatchPy patterns.
Some patterns could not be converted because they contain features that MatchPy does not support yet.
Those features include negated subpatterns (\eg \mintinline{python}{not atom<'(' [any] ')'>})
and subpatterns that allow an arbitrary number of repetitions (\eg \mintinline{python}{any (',' any)+}).

Furthermore, some of the AST patterns contain alternative or optional subpatterns, \eg \mintinline[breaklines]{python}{power<'input' args=trailer<'(' [any] ')'>>}. \fi
While these features are not directly supported by MatchPy either, they can be replicated by using multiple patterns.
For those \texttt{lib2to3} patterns, all combinations of the alternatives were generated and added as individual patterns.
This resulted in about 1200 MatchPy patterns to cover the original 36 \texttt{lib2to3} patterns.
As shown by our experiments, many-to-one matching can mostly compensate for this substantial increase in the number of patterns.
Note that the \texttt{lib2to3} patterns do not use some features of MatchPy, \eg associativity or commutativity.

For the experiments, we use 613 examples from the unittests of \texttt{lib2to3} with a total of about 900 non-empty lines of Python code.
The original \texttt{lib2to3} matcher using the set of 36 patterns is compared with MatchPy's matchers using the 1200 patterns.
A total of about 560 matches are found.

\begin{figure}[h]
    \centering
    \begin{tikzpicture}
        \begin{axis}[
            ybar,
            height=4cm,
            width=\columnwidth,
            ylabel={Time [ms]},
            symbolic x coords={codegen,many-to-one,lib2to3},
            xtick=data,
            nodes near coords,
            ymin=0,
            ymax=110,
            bar width=1.4cm,
            x=1.6cm,
            enlarge x limits={abs=0.8cm},
            xticklabel style = {font=\small},
        ]
            \addplot[fill=rwthblue] plot coordinates {
                (codegen,       18.85)
                (many-to-one,   26.12)
                (lib2to3,       86.75)
            };
        \end{axis}
    \end{tikzpicture}
    \vspace{-\baselineskip}
    \caption{Total times for \texttt{AST}}\label{fig:prop-times}
\end{figure}
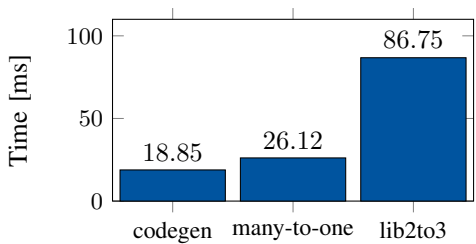

The total times for matching all subjects are shown in Figure \ref{fig:prop-times}.
One-to-one matching takes a total of more than five seconds to match and when parallelized the time goes up to 26s.
Therefore we leave these times out of the plot as they are worse by several magnitudes and we want to focus on the comparison of the other three cases.
The fastest method is the generated code with a speedup of about 4.7 over the \texttt{lib2to3} matcher.
Many-to-one matching is also about 3.4 times faster than the original implementation.
Overall, the many-to-one methods are about 200 times faster than the one-to-one matching.
This is largely caused by the significant overlap of most of the patterns which were generated from the alternatives in the original patterns.
Furthermore, less than 1\% of the patterns match any given subject.

\subsection{Rubi}

Rubi is a rule-based symbolic integrator that uses more than 6000 replacement rules \cite{Rich2009}.
It significantly outperforms Mathematica and Maple in terms of the quality of the integration.
The rules are available as Mathematica code. However, as part of a project to integrate those rules into SymPy, they  are currently being ported to MatchPy.\footnote{We would like to thank Arihant Parsoya, Abdullah Javed Nesar and Francesco Bonazzi, who have been working on integrating the Rubi rules into SymPy as part of the Google Summer of Code 2017.} We use a subset of these rules for our experiments.
A very basic example of such a rule is $x^m \rightarrow \frac{1}{m + 1} x^{m + 1}$ with the additional constraints that $x$ does not appear in $m$ and $m + 1 \neq 0$.
We use about 100 of the problems from Section 1.2 of the Rubi integration test suite as subjects.

\pgfplotstableread[col sep=comma]{rubi_results_flipped.csv}\rubiraw
\pgfplotstablesort[sort key=one-to-one,sort cmp=float <]\rubiresults{\rubiraw}

\begin{figure}[h]
    \centering
    \begin{tikzpicture}
        \begin{axis}[
            height=7cm,
            width=\columnwidth,
            xlabel={Subjects},
            ylabel={Time [ms]},
            grid=major,
            xtick=\empty,
            ymode=log,
            log origin=infty,
            legend style={at={(0.5,-0.1)},anchor=north},
            legend columns=4,
            xmin=-5,
            xmax=100,
        ]
            \addplot[fill=rwthblue,draw=rwthblue,ultra thick,only marks,mark options={scale=0.5}]
                table[
                    x expr=\coordindex,
                    y=one-to-one,
                ] {\rubiresults};
            \addlegendentry{One-to-one};
            \addplot[fill=rwthturquoise,draw=rwthturquoise,ultra thick,only marks,mark options={scale=0.5}]
                table[
                    x expr=\coordindex,
                    y=many-to-one,
                ] {\rubiresults};
            \addlegendentry{Many-to-one};
            \addplot[fill=rwthorange,draw=rwthorange,ultra thick,only marks,mark options={scale=0.5}]
                table[
                    x expr=\coordindex,
                    y=codegen,
                ] {\rubiresults};
            \addlegendentry{Codegen};
        \end{axis}
    \end{tikzpicture}
    \caption{Times for \texttt{Rubi}}\label{fig:rubi-times}
\end{figure}

The resulting times are displayed in Figure \ref{fig:rubi-times} sorted by the time that the one-to-one matching takes.
Here the speedups are less uniform than for the other applications.
The longer the overall matching takes, the smaller is the difference between the different methods.
The reason is that there is a significant overhead not related to pattern matching itself.
There are some expected correlations: The more steps (\ie rule applications) the integration takes, the longer the overall time is. The larger the term resulting from the integration is, the longer the integration takes on average.
Overall, on average the many-to-one methods are about 50\% faster than one-to-one matching.
For the fastest third of subjects (with regards to one-to-one matching time) the speedup is significantly higher with about 5.6 for code generation and 2.2 for many-to-one matching on average.

\subsection{Logic}

We use the \texttt{Prop} example from \cite{Kirchner2001} and add more examples to benchmark the performance of pattern matching on a small pattern set.
The example uses a TRS with ten rules to convert boolean formulas to algebraic normal form (ANF), \ie formulas that only contain $\wedge$ and $\oplus$.
Starting from either $\top$ or $\bot$, we generate more complicated versions of these tautologies and contradictions by applying the simplifications backwards.
For example, instead of $t$ we can write the equivalent $\neg \neg t$ or $t \wedge \bot$ for any formula $t$.
One characteristic of this problem is that formulas can become very large in the process of simplification.


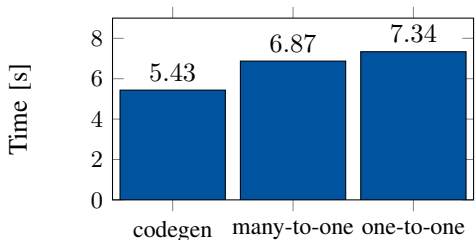
\begin{figure}[h]
    \centering
    \begin{tikzpicture}
        \begin{axis}[
            ybar,
            height=4cm,
            width=\columnwidth,
            ylabel={Time [s]},
            symbolic x coords={codegen,many-to-one,one-to-one},
            xtick=data,
            nodes near coords,
            ymin=0,
            ymax=9,
            bar width=1.4cm,
            x=1.6cm,
            enlarge x limits={abs=0.8cm},
            xticklabel style = {font=\small},
        ]
            \addplot[fill=rwthblue] plot coordinates {
                (codegen,5.434954)
                (many-to-one,6.871983)
                (one-to-one,7.336512)
            };
        \end{axis}
    \end{tikzpicture}
    \vspace{-\baselineskip}
    \caption{Average times for \texttt{Prop}}\label{fig:prop-times}
\end{figure}

The resulting times for the logic benchmark are displayed in Figure \ref{fig:prop-times}.
We only look at averages because the speedup is very uniform over all subjects.
Because there is not a lot of overlap between the patterns and the number of patterns is small, the speedup of many-to-one matching and code generation is relatively small with 7\% and 35\%, respectively.
In this case, it is also easy to write Python code by hand that performs the conversion to ANF.
Our hand-coded solution is more than three orders of magnitude faster than the one-to-one matching approach.
Therefore, we conclude that in cases with few patterns and little overlap of patterns, pattern matching is likely not a good choice in terms of performance.
It might still be interesting in terms of productivity, though, because the TRS is significantly shorter and more readable than the hand-coded solution.

\section{Productivity}

Pattern matching and term rewriting systems provide a way to solve many problems in an intuitive and expressive way.
In this section, we give an introduction of how MatchPy can be used.

MatchPy can be installed using \texttt{pip} and all necessary classes can be imported from the top-level module \texttt{matchpy}.
Expressions in MatchPy consist of constant symbols, operations, and wildcards (\ie variables).
We use Mathematica's notation\footnote{See \url{https://reference.wolfram.com/language/guide/Patterns.html}} for wildcards, \ie we append underscores to wildcard names to distinguish them from symbols.

\ifminted
MatchPy can be used with native Python types such as \texttt{list} and \texttt{int}.
The following is an example of how the subject \texttt{[0, 1]} can be matched against the pattern \texttt{[x\_, 1]}.
The expected match here is the replacement \texttt{0} for \texttt{x\_}.
Because some patterns can have multiple distinct matches, \texttt{match} is a generator that yields any match found.
Hence, we use \texttt{next} because we only want to use the first (and in this case only) match of the pattern:
\begin{minted}{pycon}
>>> x_ = Wildcard.dot('x')
>>> next(match([0, 1], Pattern([x_, 1])))
{'x': 0}
\end{minted}
\fi

\ifminted
To illustrate how powerful pattern matching is, we show how a single rewrite rule with sequence variables and constraints can be used to implement a sorting algorithm for lists.
The \texttt{CustomConstraint} class can be used to create a constraint that checks whether \texttt{a} is smaller than \texttt{b}:
\begin{minted}{python}
a_lt_b = CustomConstraint(lambda a, b: a < b)
\end{minted}
The lambda function gets called with the variable substitutions based on their name.
The order of arguments is not important and it is possible to only use a subset of the variables in the pattern.
With this constraint, we can define a replacement rule that basically describes bubble sort:
\begin{minted}{pycon}
>>> pattern = Pattern([h___, b_, a_, t___], a_lt_b)
>>> rule = ReplacementRule(pattern,
...             lambda a, b, h, t: [*h, a, b, *t])
\end{minted}
This rule finds adjacent elements in the list which are in the wrong order and swaps them.
The replacement lambda gets called with all matched variables as keyword arguments and needs to return the replacement.
This replacement rule sorts a list when applied repeatedly with \texttt{replace\_all}:
\begin{minted}{pycon}
>>> replace_all([1, 4, 3, 2], [rule])
[1, 2, 3, 4]
\end{minted}
Sequence variables can also be used to match subsequences that match a constraint.
For example, we can use the this feature to find all subsequences of integers that sum up to 5.
In the following example, we use anonymous wildcards which have no name and are hence not part of the match substitution:
\begin{minted}{pycon}
>>> x_sums_to_5 = CustomConstraint(
...                         lambda x: sum(x) == 5)
>>> pattern = Pattern([___, x__, ___], x_sums_to_5)
>>> list(match([1, 2, 3, 1, 1, 2], pattern))
[{'x': (2, 3)}, {'x': (3, 1, 1)}]
\end{minted}
\fi
More examples can be found in MatchPy's documentation \cite{MatchPyDoc}.

\section{Conclusion \& Future Work}

\subsection{Conclusions%
\label{conclusions}%
}

We present MatchPy, a pattern matching library for Python with support for sequence variables and associative/commutative functions.
This library provides tools for both one-to-one and many-to-one matching.
Because non-syntactic pattern matching is NP-hard, in the worst case the pattern matching times grows exponentially with the length of the pattern.
Nonetheless, our experiments on real world examples indicate that many-to-one matching can give a significant speedup over one-to-one matching.
However, the employed discrimination nets come with a one-time construction cost which needs to be amortized to benefit from their speedup.
In previous experiments \cite{Krebber2017,thesis}, when considering typical numbers of subjects for the respective application, many-to-one matching was always faster overall.
Therefore, many-to-one matching is likely to result in a compelling speedup in practice.
The efficiency of using many-to-one matching heavily depends on the actual pattern set, \ie the degree of similarity and overlap between the patterns.

We also use the discrimination nets to generate Python code to further improve the performance.
As expected, the generated code is faster at matching, but comes with additional time cost for generating the code in the first place.
For fixed pattern sets, code generation is the best option in terms of matching performance.
However, the generated code can get large and complicated for large pattern sets to the point where it is not viable anymore.

In our experiments, the parallelization of one-to-one matching over the patterns is not beneficial.
The parallel version is slower than the serial version in almost all cases.
This is due to the GIL in Python that only allows process based parallelism within pure Python code.
As a result, exchanging data between processes requires serializing, which introduces a significant overhead.
Independent of the choice of the matching algorithm, matching can be parallelized over multiple subjects.
For example, if we have multiple logic formulas which need to be converted to ANF, each one can be normalized independently and each replacement can run in parallel.
There is still serialization overhead, but if each process takes enough time, there is parallel speedup to be gained.
Each individual subject can then also be matched with many-to-one matching.
However, how exactly this parallelization is best implemented highly depends on the application, so we do not investigate this approach in this paper.

\subsection{Future Work}

We plan on extending MatchPy with more powerful pattern matching features to make it useful for an even wider range of applications.
The greatest challenge with additional features is likely to implement them for many-to-one matching.
In the following, we discuss some possibilities for extending the library.

\subsubsection*{Additional Pattern Features}
In the future, we plan to implement similar functionality to the \texttt{Repeated}, \texttt{Sequence}, and \texttt{Alternatives} functions from Mathematica.
These provide another level of expressiveness which cannot be fully replicated with MatchPy's current feature set.
Another useful feature are context variables as described by Kutsia \cite{Kutsia2006}.
They allow matching subterms at arbitrary depths which is especially useful for structures like XML.
With context variables, MatchPy's pattern matching would be as powerful as XPath \cite{Robie2017} or CSS selectors \cite{Rivoal2017} for such structures.
Similarly, function variables which can match a function symbol would also be useful for those applications.

\subsubsection*{C Implementation}
If pattern matching is a major part of an application, its runtime can significantly impact the overall speed.
Reimplementing parts of MatchPy as a C module would likely result in a substantial speedup.
Alternatively, adapting part of the code to Cython could be another option to increase the speed \cite{Behnel2009, Wilbers2009}.
This would also open up the possibility to benefit from parallelization by circumventing the GIL.

\subsubsection*{Code Generation}
Furthermore, the code generation has room for improvement.
For commutative matching, the many-to-one matching code is still mostly used directly with information about the patterns in the form of dictionaries.
Instead, Python code could be generated for this as well to improve performance and remove the MatchPy dependency from the generated code.
Another option is to generate Cython or C code.

\bibliographystyle{IEEEtran}
\bibliography{literature.bib}

\end{document}